\documentclass[namedreferences]{nato}
\usepackage{sprrefn}

\usepackage{graphicx}
\usepackage{amsfonts}
\usepackage{times}

\newcommand{\intOX}[1]{{\int^{#1}_{0}}}
\newcommand{\GR}{{G_{\rm R}}}
\newcommand{\GI}{{G_{\infty}}}

\newcommand{\calG}[1]{{{\cal G}({#1})}}

\newcommand{\beq}{\begin{equation}}     \newcommand{\eeq}{\end{equation}}
\newcommand{\beqa}{\begin{eqnarray}}    \newcommand{\eeqa}{\end{eqnarray}}
\newcommand{\bde}{\begin{description}}  \newcommand{\ede}{\end{description}}
\newcommand{\ben}{\begin{enumerate}}    \newcommand{\een}{\end{enumerate}}

\newcommand{\kT}{{k_{\rm B}T} } 

\newcommand{\bm}[1]{\mbox{\boldmath ${#1}$}}

\newcommand{\rmd}{{\rm d}}

\newcommand{\eqn}[1]{\beq{ #1 }\eeq}

\newcommand{\dt}{{\rmd t}}

\newcommand{\inSbracket}[1]{{\left[{#1}\right]}}

\begin{document}

\begin{opening}
\title{Internal Stress as a Link Between Macroscale and Mesoscale Mechanics}
\author{Ken Sekimoto}
\institute{\small Mati\`{e}res et Syst\`{e}mes Complexes, CNRS-UMR7057, Universit\'e
  Paris 7
  \protect\footnote{{\rm Corresponding address:} { Gulliver,} CNRS-UMR7983, ESPCI, Paris, France}
}

\runningauthor{Ken Sekimoto}
\runningtitle{Internal Stress}

\begin{abstract}
The internal (or residual) stress is among the key notions to
describe the state of the systems far from equilibrium. 
Such stress is invisible on the macroscopic scale 
where the system is regarded as a black-box. 
Yet nonequilibrium macroscopic operations
allow to create and observe the internal stress.
We present in this lecture some examples of the internal 
stress and its operations. 
We describe the memory effect in some detail, the 
process in which the history of past operations is 
recalled through the relaxation of internal stress.
\end{abstract}


\end{opening}

\section{Introduction}

In this lecture note,  the {\it internal stress} is defined as  the stress
that is  maintained within a system by itself, without the aid of external
supports or constraints.\footnote{The notion of the internal stress in this
  context therefore includes   the {\it 
    residual stress} on the one hand, since the latter has more strict definition than
  the former.  
Sometimes, however, the `{internal stress}' is used just as opposed to
the stress on the surface of a material \cite{landau-mech}  on the other
hand.}  
From outside, the internal stress is, therefore, invisible by purely
mechanical means. 
But since the stress exists locally on smaller scales, it could be observed,
for example, by stress sensitive optical probes. 
Also, once the internal mechanical balance is somehow broken,
the internal stress becomes visible from outside 
In this sense, the internal stress is a concept 
bridging between macroscale and mesoscale mechanics.
The purpose of this lecture is to draw the reader's attention
to this aspects of the internal stress.

We can find the internal stress everywhere around us:  
When we tighten the belt of our clothes, the internal stress keeps contact between the belt
under tension and our body under compression.
When we make an espresso coffee, the internal stress is between the high-pressure fluids
 inside the coffee-machine and the stretched metallic frame of the machine.
If we make a cut midway into a red sweet pepper or a watermelon, they
generally change their shape, due to 
turgor pressure of plant cells which has accumulated the internal stress.
When a pressing machine in industry spreads a metallic block, the framework
of the machine  and the block develop an internal stress, i.e., under
compression in the metallic block 
 on the one hand, under tension (and shear) in the machine's framework on
the other hand. 
The {\it tensegrity} \cite{tensegrity} is a concept of self-standing
architecture based on the internal stress. 
This is formed by wires under tension, 
rods under compression and the joints articulating those elements.
The stable structure is realized by the internal stress among these them.
This idea of tensegrity is applied to analyze the mechanical constitution in
living cells \cite{ingber}.  
The actin and microtubule filaments are thought to be joined by various active or passive binding proteins or other macromolecules.
In solid clusters atoms interacts with each other by attractive and repulsive
forces, making up a network of internal stress. 
Even the perfect salt crystal (NaCl) may be regarded as a result of internal stress due to
electrostatic forces between like and unlike charge pairs.
The atmosphere is pulled by the earth through the gravitaty, while the air also pushes the earth by the hydrostatic pressure.
Any non-trivial self-sustaining system, therefore, contains the internal
stress. 

If the internal stress is found everywhere, is this concept a ``general
abstract nonsense''  ?\footnote{The ``general
abstract nonsense'' has been a word of criticism against the category theory
\cite{categoryWM}. This theory turned out to provide 
 a powerful tool, as well as a conceptual perspective,
to many field of mathematics.} 
Yes, perhaps. But this concept can provide with a useful viewpoint.
 I will try to develop this hope in the following sections:
We first describe how the internal stress is maintained on the mesoscale (\S
2). Then we describe a mesoscale {\it sensor} that uses internal stress to
realize the reliable observation under thermal fluctuations (\S 3). 
Next we discuss briefly the generation of internal stress in the context of
far-from equilibrium process (\S 4). The last issue (\S 5) is the emergence of
the internal stress as a memory effect.

I wish that this essay serves for the analysis, modeling and designing of
various far from equilibrium phenomena such as glass dynamics, plasticity, and
active media.

\section{Mesoscale description of internal stress} 

\subsection{Incompatible stress-free states of constituent modules cause the
  internal stress} 
Under an internal stress, modules that constitute the system
undergo some deformations. Let us look back some of the examples mentioned
above.  
When a pressing machine presses down a metallic block, the
latter is deformed from its natural form, and both the block and the machine
develop the stress. Within a coffee maker, the water vapor is highly compressed with
respect to the ambient pressure, and the container of the hot water/vapor
opposes this pressure. 
The red sweet pepper before the cutting contains many cells whose shape is distorted with respect to their form under no constsraints.
Inside a cell many biopolymeric filaments are deformed by other
components of the cell. In clusters or in crystals, many pairs of atoms
are not at the distance of the lowest energy for a particular pair-interaction.

Thus, on the mesoscale, the internal stress is caused by a quasi-static
compromise among the constituent modules each of which insists on its own
stress-free state\footnote{If  this static compromise becomes unstable,
  dynamical processes, such as an oscillation, explosion, chaos, can also
  occur.} \cite{BenAmar}.

\subsection{Internal stress is the circulation of momentum flux}
\begin{figure}[ht]
 \centering
 \includegraphics[scale=0.4]{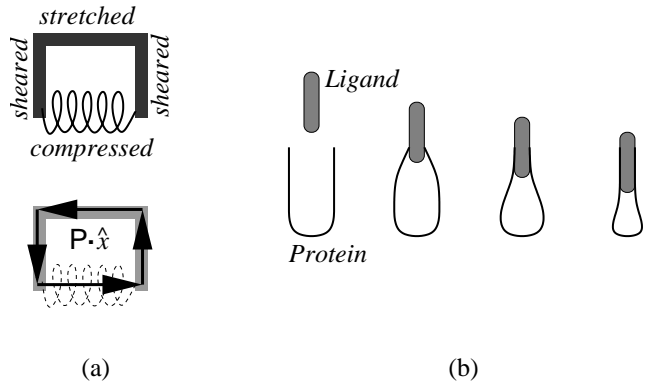} 
\caption{(a) (Top) An example of internal stress consisting of 
a spring and a framework. (Bottom) 
Schema of circulation of the flux of momentum in $x$ direction, 
${\sf P}\cdot\hat{\bm{x}}$.
 (b) ``Induced fit'' by {\it Protein} to a {\it Ligand}, 
realizing the arrival of signal to the conformational change of a receptor.
Note that the ligand is under lateral stretching, while the bottom of the
protein is under compression, as in (a).}
 \label{fig:nablaP}
 \end{figure}

The state of a system having internal stress can be represented in terms of the momentum flux.
We know that the total stress tensor $\mathbf{\sigma}$ with the minus sign carries the momentum. In (quasi-)static case of our interest 
the momentum flux density $\sf P$, therefore, writes\footnote{In general case 
 we should add the momentum flux associated with the material flow.}
\eqn{\sf P=-\mathbf{\sigma}.}
The conservation law of momentum applied to the static state, 
$\nabla\cdot {\sf P}=0,$ is nothing but the equation of static mechanical
balance of the bulk without external force, $\nabla\cdot\mathbf{\sigma}=0$. If,
for example, we focus on the flux of the $x$-component of momentum, ${\sf
  P}\cdot\hat{x}$, the conservation law is $\nabla\cdot ({\sf
  P}\cdot\hat{x})=0,$ where $\hat{x}$ is the unit vector along $x$-axis.
This is mathematically of the same form as the conservation of mass flux, 
Fig.~\ref{fig:nablaP} (a) shows an example of the permanent momentum flux.

\subsection{examples in soft materials}
\paragraph{Surface buckling of swelling gels :}
Swelling of gel can take place by many different causes:
it is sometimes enough to change the solvent \cite{Tanakabuckle87} or the
temperature \cite{DoiCargese07}. 
The swelling often develops a buckling pattern on the free surface.
While the surface region of the gel can swell freely in the direction normal 
to the surface, it is laterally constrained in order to fit with the
undeformed portion of the gel inside the bulk. 
The surface region of the gel is therefore under lateral compression.
 The buckling of the gel surface \cite{Tanakabuckle87} occurs when the
 resulting lateral compressible stress exceeds a threshold  
\cite{KSbuckling87,Suematsubuckle88}.
A similar phenomena occurs in the peripheral growth of membrane such 
 as some vegetable leaf \cite{BoudaoudCargese07}.
See also \cite{BenAmar}.
Because of the freedom perpendicular to the membrane,
the buckling pattern in this case occurs predominantly in this direction
\cite{Cerdabuckle02}.
A complementary situation can also occur: If a gel is synthesized on a
spherical substrate and pushed outward \cite{GerbalListeria}, 
the gel near the free outer surface develops a lateral tension, instead of
compressive stress. 
This stress drives the lateral fracture of gel
\cite{symbr-KS-EPJE04,jasper-sykes-pnas}, instead of buckling. 

\paragraph{Water suction at the crack tip of developing gel fracture :}
The velocity of gel's fructure is known to change if the solvent is supplied at the crack tip of developing gel fracture 
\cite{Tanakagelfracjpsj}. The mode-I opening of the crack develops the
negative pressure of the solvent within the gel.
This negative pressure pulls the menisci of the
solvent at the gel surface, which in turn compresses the gel toward inside. 
The internal stress establishes between the solvent under negative pressure
and gel under compression. Recently \cite{Trist} extrapolated the
investigation of \cite{Tanakagelfracjpsj}.  

\paragraph{Drying gel :}
A similar situation occurs at the surface of drying gel or colloidal
suspension. The surface layer of gel is under compression in 
the normal direction to the surface because it is pressed towards inside by
the surface menisci of the solvent, on the one hand, and towards outside by
the interior part of gel under osmotic pression\footnote{The capillary
  pressure $p_{\rm cap}=\gamma(K_1+K_2)$, 
where $\gamma$ is the surface tension and $\{K_1,K_2\}$ are the principal
curvatures of the meniscus, creates the permeation flow obeying the Darcy's
law, $\nabla p=-\zeta (1-\phi_{\rm gel})(\mathbf{u}_{\rm solv}-\mathbf{v}_{\rm
  gel})$, where $\zeta$ is a constant, $\phi_{\rm gel}$ is the gel's volume
fraction, and $\{\mathbf{u}_{\rm solv},\mathbf{v}_{\rm gel}\}$ are the
velocity of solvent and gel, respectively. 
The gel is pressed toward inside by the boundary condition, $\pi_{\rm
  gel}=p_{\rm cap},$ while it is pressed toward outside by the total
mechanical balance in the gel, $\nabla\pi_{\rm gel}=-\nabla p$, 
where $\pi_{\rm gel}$ is the osmotic pressure in the gel.}. 
A complementary situation occurs in ionic gels: 
The counterions between the network of polyelectrolyte gel
\cite{KhokhlovCargese07,DoiCargese07} push outward the free surface of the gel
through the electrostatic double layer on the surface (Donnan effect)
\cite{RickaTanaka}, while the network elasticity resists against the
swelling.

\begin{figure}[ht]
 \centering
 \includegraphics[scale=0.3]{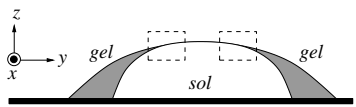} 
\caption{Schematic sectional view of a drying droplet of colloidal suspension.
The shaded region is in the gel phase, while the central part of the droplet
is still in the sol phase. In the regions marked by dashed rectancles,
the dry glassy surface layer is under compression in $x$-direction.}
 \label{fig:rides}
 \end{figure}

\paragraph{Buckling of drying colloidal suspension} 
On continuing the discussion of the drying gel, 
an important phenomenon related to the drying is the glassification of the
highly dried surface layer. 
This layer is very thin and of high elastic modulus \cite{Karine}. 
When the interior elastic part exerts lateral compressible stresses
on the glassy surface layer, it may cause 
surface buckling pattern, as is
observed before the drying fracture \cite{RideMaLangumuir}, see
Fig.~\ref{fig:rides}. \footnote{%
We should note the glassification-induced update of the stress-free state. 
Without this, the vertical compression causes
a lateral expansion as a secondary or Poisson effect. 
Given the normal compressive stress, $-\sigma_{zz}>0$, the 
induced lateral compressive stress for a laterally constrained surface
is $-\sigma_{xx}=-\sigma_{yy}=
-\frac{\nu}{1-\nu}\sigma_{zz}$ for isotropic surface, or 
$-\sigma_{xx}=-\nu -\sigma_{zz}$ and $-\sigma_{yy}=0$ when 
 the $y$ direction is not constrained \cite{RideMaLangumuir}, where $\nu$ is 
the Poisson ratio of the gel's osmotic elasticity.}

\paragraph{Permanent set of rubber crosslinking :}
Suppose we have a rubber network deformed by an external force.
If we introduce further crosslinking into this deformed network,
the network has virtually two networks whose stress free states are mutually incompatible. After removing the external force, the rubber
returns to a new apparent ``stress-free state.'' This state is called the {\it permanent set}  \cite{permanentsetFlory}.\footnote{
The purely entropic free energy of original rubber writes
$-TS_1=(\kT \nu_0/2)(\lambda_x^2+\lambda_y^2+\lambda_z^2)$,
where $\lambda_x$ etc. are the elongation ratios along $x$-direction etc. with $\lambda_x \lambda_y \lambda_z=1$. Then, the free energy, $-T(S_1+S_2)$, after the second crosslinking introduced under 
deformation, $\lambda_{x0}$ etc. with $\lambda_{x0} \lambda_{y0} \lambda_{z0}=1$ is such that 
$-TS_2=(\kT \nu_1/2)[(\lambda_x/\lambda_{x0})^2+(\lambda_y/\lambda_{y0})^2+(\lambda_z/\lambda_{z0})^2]$.
Then the permanent set has the deformation,
$\lambda_{xs}$ etc.  
obeys $\nu_0 \lambda_{xs}^2+\nu_1 (\lambda_{xs}/\lambda_{x0})^2=C$ etc, where $C$ is determined by the condition $\lambda_{xs} \lambda_{ys} \lambda_{zs}=1$ \cite{permanentsetBerry}. 
Macroscopically, the new rubber has free-energy,
$-TS_s=(\kT \nu_c/2)[(\lambda_x/\lambda_{xs})^2+(\lambda_y/\lambda_{ys})^2+(\lambda_z/\lambda_{zs})^2]$, with
$\nu_c^3=(\nu_0+\nu_1 /\lambda_{x0}^2)(\nu_0+\nu_1 /\lambda_{y0}^2)(\nu_0+\nu_1 /\lambda_{z0}^2)$. See also 
\cite{permanentsetKaang}.}\\

We thus have seen that whether or not the internal stress is visible depends on the scale of description / observation. 

\section{Sensor working on the thermally fluctuating scale} 
The sensor of signal molecules (ligands) must fulfill apparently incompatible requirements: it must respond strongly enough to the arrival of the signal through the conformational change of the sensor itself, on the one hand, but the interaction between the ligand and the sensor must be mostly neutral to reflect the density of the ligands in the environment, on the other hand. The biological receptors solved this task using the the mechanism called the {\it induced fit} \cite{koshland73}, which uses the internal stress \cite{sar}.
Fig.~\ref{fig:nablaP}(b) explains how the attraction or tension between
the ligand and the sensor deforms the latter. The analogy to the 
Fig.~\ref{fig:nablaP}(a) is evident.
It is this deformation that represents the recognition of signal, satisfying the first requirement mentioned above. In energetics term, the gain of energy by binding interaction can be compensated by the cost of deformation of the sensor. This cancellation enables the bias-free arrival and departure of the ligand, satisfying the second requirement of the sensor.

We note that the reversible detection of signal described above
has no contradiction with the theorem of Landauer and Bennett
\cite{Landauer96,bennett}, telling that we need at least $\kT\log 2$
to know the content of a single-bit memory. The latter process
decreases the entropy by $\log 2$ of the ``movable bit'', or, the observer at the compensating external work of $\kT\log 2$. In the reversible detection, the system of ligand and detector is isolated from the external observer who gains no information.

\section{Generation of the internal stress}
In many cases, the states with internal stress are metastable.
(The ionic crystal may be an exception.) To create these 
metastable states, we require far from equilibrium operations:
The pressing machine turns a motor to raise the oil/air pressure
in the piston. The coffee machine generates inner pressure by heating and boiling. Plant cells develop the pressure of vacuoles by pumps.
Colloidal gel generates the stress through evaporation flux. The permanent set of rubber network is made by chemical reaction.

The generation of internal state as metastable state is closely related to the notion of the {\it plasticity}. It is because, during the generating process, some timescales of the system is momentarily diminished at the same time that the bias is put in favor of the (future) metastable state.\footnote{We may even compare with the positive discriminations to change the society's stable state, because excessive forcing is required momentarily in both case.} 

In order to create the internal scale of a certain spatial scale, we do not necessarily need the operations on that scale.
Suppose that we bend strongly a metal rod.  
The inhomogeneity and anisotropy on the atomistic scale inside the bar can generate the metastable states of small scale, such as entangled dislocation network, a typical example of internal stress.
After removing the external forcing, the rod once yielded will show the different elasticity as well as the different yield stress \cite{TO-KS}, as compared with those of the original state. 
The new metastable state reflects the preceding far from equilibrium operation.

\section{Macroscale emergence of internal stress}
We discussed above how an external macroscale forcing
is ``internalized'' as the internal stress of  mesoscale. 
Below we describe, to some detail, the opposite case:
The internal state of mesoscale emerges as stress of macroscale through far from equilibrium processes.

\subsection{Rheological model of rubber :}
We take a simple phenomenological constitutive equation for a linear segment of rubber:
\eqn{\label{eq:constitutive}
\sigma_t=\GR \gamma_t+\GI\intOX{t} \calG{t-t'}\dot{\gamma}_{t'}\dt',}
where the tensile stress $\sigma_t$ at time $t$ consists of two terms.
The first is 
a purely entropy-elastic stress, $\GR \gamma_t$, with $\gamma_t$ being the 
elongation ($\lambda-1$) at the same time and $\GR$ being the rubber elastic modulus.
The second represents the rheological response of network chains, where
$\cal G(z)$ is the relaxation kernel satisfying  ${\cal G}(0)=1$ and ${\cal G}(\infty)=0$, and  $G_\infty$ is the lass elastic modulus.
Mathematical structure of (\ref{eq:constitutive}) is essentially the same as the 
constitutive equations of magnetic or dielectric materials.
If ${\cal G}(z)=e^{-z}$, the model reduces to the well-known Maxwell model of rheology. 
We assume that {$\gamma_t=0$ for $t<0$}.

The following rewriting of (\ref{eq:constitutive}) is physically appearing:
\eqn{\label{eq:reecrite}
\sigma_t=\GR\gamma_t+\GI \inSbracket{\intOX{t}\frac{\partial \calG{t-t'}}{\partial t'}(\gamma_t-\gamma_{t'})\dt'+\gamma_t \calG{t}}.}
If we interpret $\GI(\gamma_t-\gamma_{t'})$ as the force due to a Hooke spring
created at the time $t'$, the multiplicative factor, $\frac{\partial \calG{t-t'}}{\partial t'}\dt',$
is its survival probability until the present time $t$.
The rheological response of network chains can, therefore, be viewed as the 
permanent set which is incessantly crosslinked but also uncrosslinked with the surviving 
probability $\calG(z)$ until $z$ sec after the crosslinking.

\subsection{Internal stress in the rheological model}
Next suppose that we stopped  stretching at time $t_0$ and let loose the sample.
Shortly after releasing the stress ($t={t_0^+}\equiv t_0+\epsilon$ with small positive $\epsilon$), the stretching $\gamma_{t_0^+}$  ($0< \gamma_{t_0^+} < \gamma_{t_0} $) should obey the following equation: 
\eqn{\label{eq:release}
0=\GR \gamma_{t_0^+}+\GI\inSbracket{\intOX{t_0^+}\calG{{t_0^+}-t}\dot{\gamma}_{t'}\dt'+(\gamma_{t_0^+}-\gamma_{t_0})}.}
Since the crosslinks have finite lifetimes, the vanishing of the left hand side of   
(\ref{eq:release}) is the result of internal stress established as compromise among
the temporary crosslinks or Hooke springs. Some springs are under tension but some others should be under compression. This internal stress can remain for very long time if the temperature is below the glass temperature.

\subsection{Memory effect of rubber}
Suppose we then fix again the sample's at the relaxed length $\gamma_{t_0^+}$.
Initially $\sigma_{t_0^+}=0$ by definition. But what will be the stress $\sigma_t$ 
for $t>t_0^+$? 
We can show that
$\sigma_t$ for $t>t_0^+$ under fixed $\gamma_t=\gamma_{t_0^+}$ writes as follows:
\beqa \label{eq:remember2}
\sigma_t
&=&\GR \gamma_t+\GI\inSbracket{\intOX{t_0} \calG{t-t'}\dot{\gamma}_{t'}\dt'
+\calG{t-t_0}(\gamma_{t_0^+}-\gamma_{t_0})} \cr
&=& 
\GR[1-\calG{t-t_0}] \gamma_{t_0^+}+\GI\intOX{t_0}[
\calG{t-t'}-\calG{t-t_0}\calG{t_0-t'}]\dot{\gamma}_{t'}\dt' 
\cr &&
\eeqa
where we substituted 
 $\calG{t-t_0}\times(\mbox{\protect\ref{eq:release}})$ to go from the first line to the second.
 We note that ${\cal G}(t-t')-{\cal G}(t-t_0){\cal G}(t_0-t')$ 
in  the last integral vanishes if $\calG{z}$ is the Maxwell model, $\calG{z}=e^{-z}$.
Otherwise, a stress reappears before it returns finally to the rubber elasticity.
This phenomenon was first found experimentally \cite{miyamoto}. We 
call this reappearance of stress the memory effect. 
The Maxwell model, therefore, cannot explain the memory effect.
Intuitively, if there is more than one characteristic times in the relaxation kernel $\calG{z}$, the balance of internal stress on the mesoscale is transiently broken. And the uncompensated mesoscale stress appears as the macroscale stress. 
Due to the mathematical similarity of the constitutive equations, the memory  
effect in magnetic or dielectric systems is also understood with pertinent reinterpretation of the stress.

From information point of view, 
the {\it history} of operations in $0<t<t_0$ can be read out from
the memory effect in $t_0< t$, up to the Òmemory capacityÓ ($n$) of the system, where
 $n$ is the number of characteristic times in $\calG{z}$, i.e. $\calG{z}
=\sum^n_{j=1} a_j\, e^{-z/{\tau_j}}.$

From energetics point of view, we can assess the energy stocked in the 
state of internal stress.
We use an expression of $\sigma_t$ similar to (\ref{eq:reecrite}), which is of course equivalent to (\ref{eq:remember2}):
\eqn{\label{eq:sigmaint}
\sigma_t=\GR \gamma_{t_0^+} +\GI\inSbracket{\intOX{t_0}
\frac{\partial \calG{t_0-t'}}{\partial t'}[\gamma_{t_0^+}-\gamma_{t'}]\dt'+\calG{t}\gamma_{t_0^+}}.}
By an analogy to the potential energy of Hooke spring, the internal energy $E^{\rm int}_t$ contained by the mesoscale springs should be 
\eqn{ \label{eq:Eint}
E^{\rm int}_t\equiv \frac{\GI}{2}\intOX{t_0}[\gamma_{t_0^+}-\gamma_{t'}]^2
\frac{\partial \calG{t-t'}}{\partial t'}\dt',}
where we have ignored $\calG(t)$ term in (\ref{eq:sigmaint}).
Experimentally, the corresponding quantity has been measured using the calorimetry 
\cite{hasan}. They prepared glassy rubber samples with or without pre-stretching,
and compared the exothermic heat upon slow warming. The pre-stretched sample,
i.e. the sample containing the internal stress in the present context, released
an excess heat. 
If this heat corresponds to the decrease of $E^{\rm int}_t$ in (\ref{eq:Eint}),
the results implies that the mesoscale Hooke springs are {\it not} of entropic origin.
\\

I thank my colleagues, Y. Tanaka, T. Ooshida, N. Suematsu, K. Kawasaki and Y. Miyamoto, J. Prost, F. {J\"{u}licher}, H. Boukellal and A. Bernheim-Grosswasser,
whose collaborations are cited above. I thank M. Ben-Amar for having shown unpublished paper, and A. Daerr  for a critical comment. 
I appreciate the organizers of this School.

\bibliographystyle{sprnamed}
\bibliography{sekimoto_cargese_bib}

\end{document}